\documentclass{acm_proc_article-sp}

\usepackage{auto-pst-pdf}
\usepackage{subcaption}

\DeclareMathOperator{\KL}{D_{KL}}

\begin{document}

\title{Application of Kullback-Leibler divergence for short-term user interest detection}
\subtitle{[Extended Abstract]}

\numberofauthors{3}

\author{
\alignauthor
Maxim A. Borisyak
    \affaddr{Moscow Institute of Physics and Technology}\\
    \email{borisyak@phystech.edu}
\alignauthor
Roman V. Zykov\\
       \affaddr{Retail Rocket}\\
       \email{rzykov@retailrocket.ru}
\alignauthor
Artem E. Noskov\\
	\affaddr{Retail Rocket}\\
    \email{a.e.noskov@gmail.com}
}

\date{\today}

\maketitle
\begin{abstract}
  Classical approaches in recommender systems such as collaborative filtering are concentrated mainly on static user preference extraction.
  This approach works well as an example for music recommendations when a user behavior tends to be stable over long period of time,
  however the most common situation in e-commerce is different which requires reactive algorithms based on a short-term user activity analysis.
  This paper introduces a small mathematical framework for short-term user interest detection formulated in terms of item properties
  and its application for recommender systems enhancing.
  The framework is based on the fundamental concept of information theory --- Kullback-Leibler divergence.
\end{abstract}




\section{Introduction}
Artificial Intelligence point of view considers a recommender system as an agent with user as its environment.
Since a user is an agent itself it is naturally to assume that by using recommender systems user usually pursue some personal goals.
The most general objective of recommender systems is to respond accordingly to user behavior and so to his goals.
However the goals only partially depend on user global preferences.

When user behavior is determined mostly by global preferences (as in music) the objective degenerates correspondingly.
In e-commerce user goals are usually dictated by some external reasons unknown to the recommender system.
Another related difference is that amount of data needed for obtaining an adequate estimation of user preferences
is usually far in excess of the same amount for other more specialized areas (for example, movie recommendations).
These two factors make user behavior in e-commerce appear to be more depended on
short-term personal goals rather than on static preferences from the recommender system perspective, which justifies value
of short-term analysis.

\section{Definitions and assumptions}
All recommender systems receive an event flow from each user but we consider problem of splitting
the event flow into sessions solved.

\newdef{definition}{Definition}
\begin{definition}
User session $s$ is defined as finite sequence of items user had interaction (usually view) with pursue one particular goal:
\begin{equation}
	s = \{ i_j \mid i \in I\}_{j = 1}^m
\end{equation}
where $I$ is set of all items.
\end{definition}

We also introduce set of properties $K$ which is defined for each item:
\begin{equation}
	\forall k \in K, \forall i \in I:\ f(i, k) \in V_k
\end{equation}
where $V_k$ --- possible values of property $k$. For simplicity $V_k$ is always a finite set.

\subsection{Model of user behavior}
We consider user as an agent trying to fulfill its own purpose and
our main assumption is that user actions are dictated by his will to find an item with particular set of properties $U \subseteq K$
(for example, color, size or price).
Taking into account additional assumption about rationality of users we can regard session as a trace of some kind of optimization and
comparison process performed in the partially observed environment (items and their descriptions)
which points to the stochastic nature of the search process.
This interpretation of user behavior allows a lot of mathematical models which may perfectly fit
into the suggested method, but for the purpose of the paper
we will adhere to one of the simplest: the user session $s$ is viewed
as samples of random variable $\psi^s$ with distribution $\Psi^s$:
\begin{equation}
    i_j \sim \Psi^s,\ j = 1, \dots, m \label{eq:model}
\end{equation}

$\psi^s$ here defines real user interest within the model with regards to observation limits of recommender systems.

It should be noted that \ref{eq:model} is also a definition of user session,
however, in practice the splitting of event flow can be done well enough by setting maximal time difference between adjacent events and
by a few additional heuristics (for instance, an purchase event finalizes current session).

\section{User interest}
User interest in some property $k \in K$ is determined relatively to the common interest in $k$.
Suppose $G$ denotes general distribution of items, prior probability of item $i \in I$ appearing in an event and
$G_k$ denotes distribution of values of property $k$. Distributions $\Psi^s_k$ are defined in the similar way.

\begin{definition}
User interest within session $s$ is the set of properties $U^s$:
\begin{equation}
	U^s = \{ k \mid \Psi^s_k \neq G_k,\ k \in K \} \label{eq:interest}
\end{equation}
\end{definition}

Of course, in practice \ref{eq:interest} is hard to check directly since distribution $\Psi^s_k$ is
known only approximately\footnote{Distribution estimation error is usually quite big since common user session contains approximately 5-10 events}.
A measurement of difference between two distributions allows to
apply statistical hypothesis testing and Kullback-Leibler divergence\cite{kullback1951information}\cite{cover2012elements} is a natural choice\cite{eguchi2006interpreting} for the test statistic\footnote{As an
alternative, for example, consider Kolmogorov-Smirnov test\cite{lopes2011kolmogorov}.}.
\begin{definition}
Let $P(\omega)$ and $Q(\omega)$ denote distributions over finite space $\Omega$.
Then Kulback-Leibler relative information gain of $Q$ from $P$ is: 
\begin{equation}
	\KL(P \mid Q) = \sum_{\omega \in \Omega} \left( P_y \cdot \log \frac{P_y}{P_x}\right)(\omega)
\end{equation}
\end{definition}

Obviously, in our case:
\begin{equation}
	\Psi^s_k = G_k \Leftrightarrow \KL(\Psi^s_k \mid G_k) = 0
\end{equation}
Definition \ref{eq:model} can be reformulated correspondingly\cite{dahlhaus1996kullback}.
Now we can formulate two statistical hypothesis for each $k \in K$ corresponded to $k \in U^s$ and $k \notin U^s$:
\begin{eqnarray}
	H_0&:& \KL(\Psi^s_k \mid G_k) = 0\\
	H_1&:& \KL(\Psi^s_k \mid G_k) > 0
\end{eqnarray}
and if $\widehat{\Psi}^s_k$ denotes estimation of $\Psi^s_k$ the decision rule is following:
\begin{equation}
	\delta_k(s) = \begin{cases}
		k \in U^s & \text{if } \Delta^s_k < \varepsilon^m_k\\
		k \notin U^s& \text{otherwise}
	\end{cases}
\end{equation}
where
$\Delta^s_k = \KL\left(\widehat{\psi}^s_k \mid G_k\right)$.

Since distributions $G_k$ are known in advance, distribution of $\Delta^s_k$ under $H_0$ can be also precalculated\footnote{An important moment here is that thresholds $\varepsilon^m_k$ considerably depend on the length $m$ of the session.}.
Authors recommend to do it simply by sampling from $G_k$ since additional assumptions and modifications may require estimations of $\Psi$ different from
the empirical distribution function which may bring unnecessary complications.

It should be noted, that one of the canonical ways to obtain levels $\varepsilon_k$ is by minimizing the risk function, which
may be quite complicated because end algorithm produces sequence of action and so the risk function may involve
user-system interaction component. Since the risk function can be directly inferred from selected quality function for end algorithm,
it is much simpler to consider $\varepsilon_k$ as meta-parameters.

\section{Algorithm enhancing}
The primary aim of short-term interest detection is to enhance recommender systems.
We consider base recommender algorithm $R: I \rightarrow I^N$ defined by weight function $w(\cdot)$:
\begin{equation}
    R(i) = \mathop{\mathrm{arg\,topN}}_{j \in I, j \neq i} w(j)
\end{equation}
where $\mathop{\mathrm{arg\,topN}}$ is defined analogously to $\mathrm{arg\,max}$ operator.

Usually the enhancing by considering short-term user interest is reasonable when $R(\cdot)$ is an offline algorithm and
does not depend on whole session $s$ and the system respond only to current event $s_m$\footnotemark, however it may depend on long-term user history:
\begin{displaymath}
    R_u(s) = R_u(s_m)
\end{displaymath}
where $u$ denotes user whom session $s$ belongs to.

\footnotetext{ This restriction could be easily expanded,
for example, for algorithms that take into account sequence of  events limited by predefined length.
The general idea is that if we do not want to utilize the same information
twice the base algorithm may not widely share its sources with the enhancing algorithm.
Offline algorithms usually satisfy this requirement since it is hard to precalculate recommendations for all possible sessions.}

We demonstrate only a simple example of enhancing:
\begin{eqnarray}
    c^s(j) & = & \prod_{k \in U} \frac{\widehat{\Psi}^s_k(j)}{G_k(j)} \label{eq:interestc}\\
    R^*(s) & = & \mathop{\mathrm{arg\,topN}}_{j \in I, j \neq i} c^s(j) w(j) \label{eq:enhanced}
\end{eqnarray}
where $i$ is the last item in the session and $c^s(j)$ is the interest coefficient in the item~$j$.

In a very simple case when $w(j) = G(j)$ $c^s(j) w(j)$ corresponds to
estimation of posterior\footnotemark probability of item $j$ given session $s$ under our model of user behavior.

Expression for $c^s(j)$ and $R^*(s)$ should be adopted for the features of $R(\cdot)$ once the nature of the weights becomes more specific.
The expressions \ref{eq:interestc} and \ref{eq:enhanced} reflect probabilistic nature of $w(\cdot)$ when
recommendations are based on prior probabilities which then are rescaled to posterior given session $s$ as the evidence.

\footnotetext{If all properties are considered to be independent.}

\section{Experiment}
For the experiment the following model was used:
\begin{equation}
	\widehat{\Psi}^s_k(v) = \left(1 - e^{\alpha_k |s|} \right) f^s_k(v) + e^{\alpha_k |s|} G_k(v) \label{eq:psi_estimation}
\end{equation}
where $f^s_k(v)$ is frequency of value $v$ in session $s$, $\alpha_k$ are considered as meta-parameters.
The additional smoothing is applied in order to bring computational stability and to avoid low-frequency problem.
It should be noted that the optimal $\alpha_k$ are considerably greater than zero ($\approx 0.5$) for our evaluation.

The best available proprietary algorithm, cosine similarity by statical features, was used as base algorithm.
Enhancing was performed by \ref{eq:interestc} and \ref{eq:interestc}.
To demonstrate importance of short-term user interest detection we included two simple algorithms for enhancing.
\begin{eqnarray}
    w_{\text{static}}(j) &=& \cos(f(i), f(j))\\
	w_1(j) &=& 1\\
	w_{\text{popular}}(j) &=& G(j)
\end{eqnarray}

Data for the experiment was collected from a e-commerce website specialized on appliances and gadgets.
This category has very rich descriptions (properties) for each item and is perfectly suitable for the suggested algorithm in general.

A simplified version of DCG metric and simple 'hit' metric were used as quality functions.
Each user session $s$ ($m = |s|$) was divided into two parts:
\begin{itemize}
  \item history: $h = [s_1, \dots, s_{m - 1}]$
  \item validation: $t = s_m$
\end{itemize}

Let $r_l$ denote recommendation of rank $l$ for session $h$.
In this terms the evaluation metrics can be expressed as:
\begin{eqnarray}
	\mathrm{DCG}(N) &=& \sum_{l = 1}^N \frac{\mathrm{rel}\,r_l }{\log_2 (l + 1) } \\
	\mathrm{Id}(N) &=& \sum_{l = 1}^N \mathrm{rel}\,r_l
\end{eqnarray}
where
\begin{displaymath}
  \mathrm{rel}\,x = \begin{cases} 1 & \text{if } t = x\\ 0 & \text{otherwise} \end{cases}
\end{displaymath}

\begin{figure*}[b]
\centering
	\epsfig{file=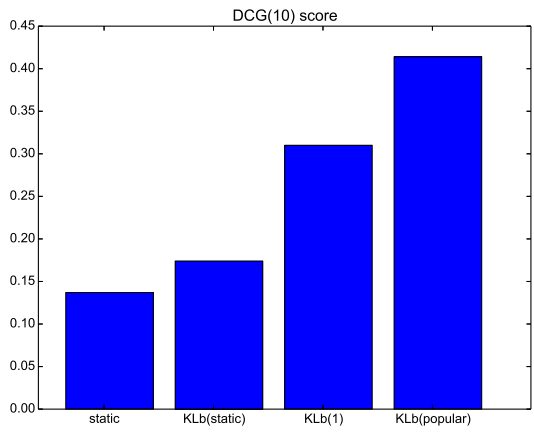, width=8.45cm}
	\centering
	\epsfig{file=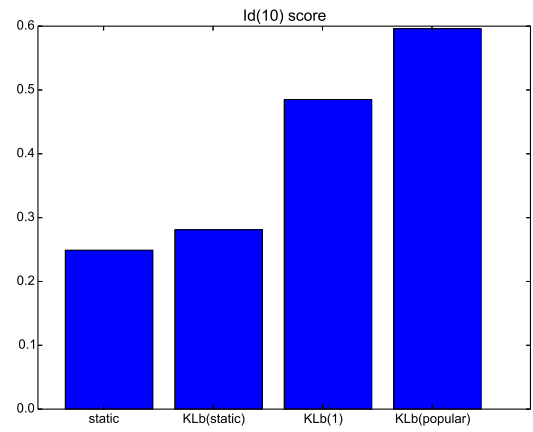, width=8.45cm}
\caption{Results of the experiment. 'static' denotes original base algorithm, KLb($\cdot$) denotes enhanced algorithm.}
\label{fig:results}
\end{figure*}

The experiment results are show on figure \ref{fig:results}.

\section{Acknowledgments}
The authors would like to thank \textit{Retail Rocket} for supporting this research.

\bibliographystyle{abbrv}
\bibliography{paper}

\end{document}